# Superconducting Symmetry Studied from Impurity Effects in Single-Crystal Fe$_{1-y}$M$_y$Se$_{0.3}$Te$_{0.7}$ (M = Co, Ni, Zn)

Takuya Inabe[1], Takayuki Kawamata[1,*], Takashi Noji[1], Tadashi Adachi[1], and Yoji Koike[1]

[1]Department of Applied Physics, Tohoku University, 6-6-05 Aoba, Aramaki, Aoba-ku, Sendai 980-8579, Japan

We have investigated the suppression of the superconducting transition temperature, $T_c$, with an increase of the residual resistivity, $\rho_0$, through the substitution of M (M = Co, Ni, Zn) for Fe in Fe$_{1-y}$M$_y$Se$_{0.3}$Te$_{0.7}$ single crystals, in order to clarify the symmetry of the superconducting gap in FeSe$_{1-x}$Te$_x$. Small, large and very small suppression of $T_c$ have been observed through the Co, Ni and Zn substitution, respectively. The magnitude of the suppression rate is hardly explained in terms of the pair-breaking effect due to potential scattering calculated based on the Abrikosov-Gor'kov theory, even if errors in the estimation of the carrier concentration, effective mass and $\rho_0$ were taken into account. Accordingly, these results suggest that the superconducting symmetry is the S$_{++}$-wave in FeSe$_{1-x}$Te$_x$.

# 1. Introduction

Since the discovery of the Fe-based superconductor LaFeAsO$_{1-x}$F$_x$,[1] many kinds of Fe-based superconductors with two-dimensional network of Fe-pnictide or Fe-chalcogenide tetrahedra have been found. Since the superconducting transition temperature, $T_c$, exceeds 50 K in carrier-doped REFeAsO (RE = Pr, Sm and Nd) systems,[2-5] the Cooper pairs are expected to be formed not simply by the ordinary electron-phonon interaction but by an unconventional one. The Fe-based superconductors have two kinds of Fermi surfaces at the Γ and M points in reciprocal space.[6-9] The Fermi surfaces are formed by five orbitals due to Fe $3d$ electrons playing an important role in the appearance of superconductivity. At present, it is controversial whether signs of the superconducting gap of these Fermi surfaces at the Γ and M points are reverse or not. In the case of the sign-reversal, namely, the so-called S$_\pm$-wave symmetry, the spin fluctuation due to the nesting between the Fermi surfaces at the Γ and M points is suggested to be relevant to the pairing mechanism.[10,11] In the case of the same sign, namely, the so-called S$_{++}$-wave symmetry, on the other hand, the orbital fluctuation between five orbitals of Fe $3d$ electrons is suggested to be relevant.[12,13] From the early time of the discovery of the Fe-based superconductors, many experimental results have been understood without contradiction by considering the S$_\pm$-wave symmetry. However, almost all studies of the impurity effect on the superconductivity have revealed that the rate of the suppression of $T_c$ is too small to be explained as being due to the pair-breaking effect in a superconductor with the S$_\pm$-wave symmetry,[14-18] because large suppression of $T_c$ due to the scattering by nonmagnetic impurities is expected to take place in a S$_\pm$-wave superconductor, being calculated based on the Abrikosov-Gor'kov (AG) theory.[19]

The superconductor FeSe$_{1-x}$Te$_x$ is suitable for the study of the mechanism of superconductivity, for it is the simplest in the crystal structure among the Fe-based superconductors. Experimental results of the specific heat,[20,21] thermal conductivity,[22] and penetration depth estimated from muon spin relaxation[23] have suggested that FeSe$_{1-x}$Te$_x$ is a superconductor with a nodeless gap. Furthermore, studies of the coherence factor by scanning tunneling microscopy/spectroscopy (STM/STS)[24] and microwave conductivity[25,26] in FeSe$_{1-x}$Te$_x$ have suggested the presence of the S$_\pm$-wave symmetry. Experimental results of the impurity effect in Fe$_{1-y}$Co$_y$Se$_{0.4}$Te$_{0.6}$ have also exhibited to be consistent with the S$_\pm$-wave symmetry,[27] because the $T_c$ suppression is not contradictory to the calculated one based on the AG theory. However, there remains ambiguity in the estimation of the rate of the $T_c$ suppression based on the AG theory. For example, errors in the estimation of the carrier concentration, $n$, and the effective mass, $m^*$, must be taken into account in the use of the AG theory.

Recently, we have found that bulk superconductivity appears through the annealing in vacuum (~ 10$^{-6}$ Torr) in a wide range of $x$ in FeSe$_{1-x}$Te$_x$ and that the annealed crystal of $x$ = 0.7 exhibits the strongest and homogeneous superconductivity in FeSe$_{1-x}$Te$_x$.[21,28] So, we have investigated the suppression of $T_c$ through the substitution of M (M = Co, Ni, Zn) for Fe in Fe$_{1-y}$M$_y$Se$_{0.3}$Te$_{0.7}$ single crystals, in order to clarify the symmetry of the superconducting gap. The characterization of the obtained single-crystals by powder x-ray diffraction and inductively coupled plasma optical emission spectrometry (ICP-OES) is described in Sect. 3.1. Impurity effects on the electronic state in Fe$_{1-y}$M$_y$Se$_{0.3}$Te$_{0.7}$ studied by the magnetic susceptibility, $\chi$, and the Hall coefficient, $R_H$, are described in Sect. 3.2. Impurity effects on the superconductivity and the residual resistivity, $\rho_0$, are shown in Sect. 3.3 and then the symmetry of the superconducting gap is discussed in Sect. 3.4.

## 2. Experimental

Single crystals of $Fe_{1-y}M_ySe_{0.3}Te_{0.7}$ (M = Co, Ni, Zn) were grown by the Bridgman method.[28] Raw materials of Fe, Co, Ni, Zn, Se, and Te prescribed in the nominal composition (3 % more Fe was added taking into account the presence of excess Fe) were thoroughly mixed in an Ar-filled glove box and sealed in an evacuated quartz tube. Since the quartz tube often cracked upon cooling, the tube was sealed in another large-sized evacuated quartz tube. The doubly sealed quartz ampoule was placed in a furnace so that single crystals were grown using the temperature gradient in the furnace. The ampoule was heated at 600 ºC for 100 h and successively at 950 - 1050 ºC for 30 h, and then cooled down to 650 ºC at the rate of 2 - 3 ºC/h, followed by furnace cooling down to room temperature. As-grown crystals obtained thus were annealed at 400 ºC for 200 h in vacuum (~ $10^{-6}$ Torr).

Grown crystals were characterized by powder x-ray diffraction. The chemical composition was confirmed by ICP-OES. The $\chi$ was measured using a superconducting quantum interference device (SQUID) magnetometer (Quantum Design, MPMS). Measurements of the electrical resistivity along the $ab$-plane, $\rho_{ab}$, and $R_H$ were carried out by the standard dc four-probe method using a commercial apparatus (Quantum Design, PPMS). Lead wires of gold were attached to each side surface parallel to the $c$-axis of a rectangular single-crystal (3.5 × 2 × 0.2 mm$^3$) with silver paste (DuPont, 4922) so that the contact resistance between the single crystal and lead wires was less than 1 Ω. The $R_H$ was measured in magnetic fields parallel to the $c$-axis up to 9 T.

## 3. Results and Discussion

### 3.1 Characterization of single crystals

Figure 1 shows the $y$ dependence of the lattice parameters $a$ and $c$ of $Fe_{1-y}M_ySe_{0.3}Te_{0.7}$ (M = Co, Ni, Zn) single crystals. It is found that $a$ hardly changes, while $c$ monotonically changes, as observed in the previous works.[27,29] This indicates that the substitution of M for Fe is realized.

Table I lists the chemical compositions determined by ICP-OES of the obtained as-grown single-crystals of $Fe_{1-y}M_ySe_{0.3}Te_{0.7}$ (M = Co, Ni). It is found that each chemical composition almost coincides with the nominal composition, indicating that the substitution is successful. Here, it is noted that the content of Fe is a little larger than the nominal one. In $FeSe_{1-x}Te_x$, it has been reported that the excess Fe induces an upturn of $\rho_{ab}$ at low temperatures,[30,31] leading to the suppression of bulk superconductivity. It has been reported that the content of the excess Fe decreases through the annealing in oxygen for a short time.[32,33] Accordingly, it is likely that the content of the excess Fe in the present single-crystals decreases through the annealing in vacuum ($\sim 10^{-6}$ Torr), leading to the appearance of bulk superconductivity in $FeSe_{0.3}Te_{0.7}$.[21,28] Since the content of the excess Fe in as-grown single-crystals of $Fe_{1-y}M_ySe_{0.3}Te_{0.7}$ is not dependent on $y$ so much, it is expected that the $y$ dependence of $T_c$ in the present annealed single-crystals of $Fe_{1-y}M_ySe_{0.3}Te_{0.7}$ is not so influenced by the excess Fe.

### 3.2 Impurity effects on the electronic state

Figure 2 shows the temperature dependence of $\chi$ in a magnetic field of 1 T parallel to the $c$-axis for $FeSe_{0.3}Te_{0.7}$ and $Fe_{0.95}M_{0.05}Se_{0.3}Te_{0.7}$ (M = Co, Ni, Zn) single crystals. It is found that a little Curie-like upturn is observed at low temperatures for all the crystals.

However, there is no increase of the Curie-like upturn through the Co, Ni and Zn substitution, indicating that both Co, Ni and Zn ions behave as not magnetic but nonmagnetic impurities. Additionally, it is noted that a kink observed around 120 K in every crystal is due to a very small amount of $Fe_3O_4$ contained as impurities and that the little Curie-like upturn at low temperatures will also be due to a very small amount of impurities.

Figure 3 shows the temperature dependence of $R_H$ of $FeSe_{0.3}Te_{0.7}$ and $Fe_{0.95}M_{0.05}Se_{0.3}Te_{0.7}$ (M = Co, Ni, Zn) single crystals. The sign of $R_H$ is positive in $FeSe_{0.3}Te_{0.7}$, indicating that holes are dominant in the conduction of $FeSe_{0.3}Te_{0.7}$. At low temperatures, $R_H$ increases with decreasing temperature in $FeSe_{0.3}Te_{0.7}$. These are consistent with the results in the previous works.[34,35] The upturn of $R_H$ at low temperatures is suppressed a little, in some degree and markedly through the Zn, Co and Ni, respectively. Since the upturn of $R_H$ is known to be induced by antiferromagnetic spin fluctuations[36-38] or orbital fluctuations[38,39], these results suggest that the fluctuations change in greater or less degree through the substitution of M for Fe, owing to the change of the Fermi surfaces. That is, since Ni ions supply the conduction bands with twice as many electrons as Co ions, it is understood that the Ni substitution changes the Fermi surfaces, spin or orbital fluctuations and $R_H$ at low temperatures more markedly than the Co substitution. On the other hand, the Zn substitution is understood not to change $R_H$ so much, because Zn 3$d$ electrons are localized around Zn ions on account of their energy levels of 3$d$ electrons much deeper than Fe 3$d$ electrons, so that Zn ions do not supply the conduction bands of Fe with electrons so much. These interpretations are supported by angle-resolved photoemission spectroscopy (ARPES) results of the present single-crystals obtained by Sudayama *et al.*[40] that the Fermi level shows a rigid-band shift through the Co and Ni substitution while it does not through the Zn substitution. These interpretations are also consistent with the above results of $\chi$ indicating that both Co, Ni and Zn ions behave as nonmagnetic impurities.

*3.3 Impurity effects on the superconductivity and residual resistivity*

Figure 4 shows the temperature dependence of $\chi$ in a magnetic field of 10 G parallel to the *c*-axis on warming after zero-field cooling for $Fe_{1-y}M_ySe_{0.3}Te_{0.7}$ (M = Co, Ni, Zn) single crystals. It is found that $T_c$, defined as the intersecting point between the extrapolated line of the steepest part of the Meissner diamagnetism and zero susceptibility, is suppressed through the Co and Ni substitution, while $T_c$ is not suppressed through the Zn substitution.

Figure 5 shows the temperature dependence of $\rho_{ab}$ of several pieces of $Fe_{1-y}M_ySe_{0.3}Te_{0.7}$ (M = Co, Ni, Zn) single crystals. Values of $T_c$, defined as the midpoint of the superconducting transition curve in the $\rho_{ab}$ vs. *T* plot, are shown in Fig. 6, together with those estimated from the $\chi$ measurements described above. Although values of $T_c$ estimated from $\rho_{ab}$ are a little different piece by piece, it is found that both values of $T_c$ estimated from $\rho_{ab}$ and $\chi$ are in rough agreement with each other and that the suppression of $T_c$ is large, small and very small through the Ni, Co and Zn substitution, respectively.

Figure 5 also reveals that $\rho_{ab}$ in the normal state tends to increase with increasing *y*. However, there is ambiguity in the estimation of $\rho_0$, because values of $\rho_{ab}$ in the normal state are a little different piece by piece and moreover because $\rho_{ab}$ of each M-substituted single-crystal exhibits an upturn at low temperatures due to the weak localization effect caused by the substituted M ions. Therefore, we have defined the maximum value of $\rho_0$, $\rho_0^{\mathrm{max}}$, as the maximum value of $\rho_{ab}$ at low temperatures and defined the minimum value of $\rho_0$, $\rho_0^{\mathrm{min}}$, as the value extrapolated to 0 K of the $T$-linear part of $\rho_{ab}$ at low temperatures, as shown in insets of Fig. 5.

*3.4 Symmetry of the superconducting gap*

On the basis of the present results, we discuss the pair-breaking effect of nonmagnetic impurities using values of $T_c$, $\rho_0^{\mathrm{min}}$ and $\rho_0^{\mathrm{max}}$ estimated from $\rho_{ab}$ measurements, because both substituted Co, Ni and Zn behave as nonmagnetic impurities as mentioned above. In the case of the $S_{++}$-wave symmetry, $T_c$ is expected not to be suppressed so much. In the case of the $S_{\pm}$-wave symmetry, on the other hand, $T_c$ is expected to be markedly suppressed on account of the pair-breaking effect due to potential scattering by substituted nonmagnetic impurities, and the rate of the $T_c$ suppression is given by the following equation based on the AG theory,[19]

$$\ln\left(\frac{T_{c0}}{T_c}\right) = \psi\left(\frac{1}{2}+\frac{\alpha}{2t}\right) - \psi\left(\frac{1}{2}\right), \quad (1)$$

where $\psi(z)$ is the digamma function defined as $\psi(z) \equiv \ln[d\Gamma(z)/dz/\Gamma(z)]$, $T_{c0}$ is $T_c$ of the non-substituted crystal, and $t = T_c/T_{c0}$. The pair-breaking parameter $\alpha \equiv \hbar/(2\pi k_B T_{c0}\tau)$ depends on the scattering relaxation time of carriers, $\tau$, which is estimated from $\rho_0$. Here, $\hbar$ is the Plank constant and $k_B$ is the Boltzmann constant. In the case of $T_c \approx T_{c0}$ and $\alpha/2t \ll 1/2$, using the relation $\rho_0 = m^*/(ne^2\tau)$, Eq. (1) is approximately given by

$$T_c = T_{c0} - \frac{\pi\hbar ne^2}{8k_B m^*}\rho_0, \quad (2)$$

where $e$ is the elementary electric charge. That is, the rate of the $T_c$ suppression is proportional to only the magnitude of $\rho_0$ at $T_c \approx T_{c0}$.

Figure 7 shows the variation of $T_c$ with $\rho_0^{\mathrm{min}}$ and $\rho_0^{\mathrm{max}}$ of $Fe_{1-y}M_y Se_{0.3}Te_{0.7}$ (M = Co,

Ni, Zn) single crystals. It is found that $T_c$ is suppressed markedly, in some degree and a little through the Ni, Co and Zn substitution, respectively, with increasing $\rho_0^{min}$ and $\rho_0^{max}$. Shadow areas are expected ones in a $S_\pm$-wave superconductor calculated using Eq. (2) and taking into account errors in the estimation of $n$ and $m^*$. Here, $n$ has been estimated from the value of $R_H$ in FeSe$_{0.3}$Te$_{0.7}$ as $2 - 3 \times 10^{21}$ cm$^{-3}$, which is consistent with the result in the previous work.[34] On the other hand, $m^*$ has been estimated as 6 - 20 $m$ ($m$ : free-electron mass) according to the result of ARPES.[41] It is found that the $T_c$ suppression is weaker than the expected one in a $S_\pm$-wave superconductor, though the $T_c$ suppression of Ni-substituted crystals looks comparable with the expected one in Fig. 7 (a). Here, it is not reasonable to regard only the $T_c$ suppression of Ni-substituted crystals as being explained based on the AG theory, as follows. That is, Eq. (2) means that the rate of the $T_c$ suppression is independent of the kind of impurities but dependent on only $\rho_0$, namely, the probability of the potential scattering. Therefore, if only the potential scattering suppressed the superconductivity in a $S_\pm$-wave superconductor, variations of $T_c$ with $\rho_0^{min}$ or $\rho_0^{max}$ in Co-, Ni- and Zn-substituted crystals should be located on the same line in Fig. 7. Accordingly, it is reasonable to regard the $T_c$ suppression in Co- and Ni-substituted crystals as being due to the electron doping by Co and Ni ions as mentioned in Sect. 3.2 in addition to the potential scattering. The reason is as follows. When electrons are doped, the Fermi surface of holes at the $\Gamma$ point shrinks, while the Fermi surface of electrons at the M point is enhanced, so that the nesting between the Fermi surfaces becomes worse. Therefore, the superconductivity is suppressed through the electron doping in the case that the spin fluctuation enhanced by the nesting is relevant to the pairing. Even in the case that the orbital fluctuation is relevant to the pairing, it is possible that the superconductivity is suppressed through the electron doping due to the change of the contributions of orbitals to the superconductivity. It is also understood that the rate of the $T_c$ suppression through the Ni substitution is more marked than through the Co substitution, because Ni ions supply the conduction bands with twice as many electrons as Co ions. The $T_c$ suppression due to carrier doping has been observed in LaFe$_{1-y}$M$_y$AsO$_{0.89}$F$_{0.11}$ (M = Co, Ni)[14,16] and BaFe$_{2-y}$M$_y$As$_2$ (M = Co, Ni, Cu)[42] also. As for Zn-substituted crystals, on the other hand, the $T_c$ suppression is weaker than in Co- and Ni-substituted crystals, for Zn ions do not operate to supply the conduction bands with electrons so much, as mentioned in Sect. 3.2. Therefore, it is concluded that the $T_c$ suppression due to the potential scattering by Zn is much weaker than the expected one in a $S_\pm$-wave superconductor. In the long run, these results suggest the presence of the $S_{++}$-wave symmetry in FeSe$_{0.3}$Te$_{0.7}$.

As mentioned in Sect. 3, the previous work of impurity effects in Fe$_{1-y}$Co$_y$Se$_{0.4}$Te$_{0.6}$ by Nabeshima et al.[27] has supported the presence of the $S_\pm$-wave symmetry, based on the result that the $T_c$ suppression through the Co substitution is not contradictory to the calculated one using the AG theory. A difference between the previous and present works is that the range of values of $\rho_0$ and $T_c$ in the previous work is much narrower than that in the present work. As a result, it is possible to regard their values of $\rho_0$ and $T_c$ as being explained in terms of the $S_\pm$-wave symmetry within the experimental accuracy. However, the present results cannot be explained in terms of the $S_\pm$-wave symmetry, even if the experimental accuracy were taken into account. Another difference between the previous and present works is that both the increase in $\rho_0$ and the decrease in $T_c$ through the Co substitution in the previous work are much smaller than those in the present work, though the reason is not clarified.

As mentioned in Sect. 1, the STM/STS study has suggested the presence of the $S_\pm$-wave symmetry,[24] which is different from the present conclusion. According to the recent theory by Efremov *et al.*,[43] it is possible that a $S_\pm$-wave superconductor changes to a $S_{++}$-wave one through the substitution of impurities. This may explain both the STM/STS and present results well, but further study is necessary to be conclusive.

## 4. Summary

We have investigated the suppression of $T_c$ and the increase in $\rho_0$ through the substitution of M (M = Co, Ni, Zn) for Fe in order to clarify the symmetry of the superconducting gap, using $Fe_{1-y}M_y Se_{0.3}Te_{0.7}$ single crystals of high quality annealed in vacuum (~ $10^{-6}$ Torr) at 400 °C for 200 h. First, it has been found that both Co, Ni and Zn ions behave as nonmagnetic impurities, because no increase of the Curie-like upturn of $\chi$ was observed through the substitution of M. Next, it has been found from $R_H$ measurements that Co and Ni ions supply the conduction bands with one and two electrons, respectively, while a Zn ion does not. Finally, it has been found that rates of the $T_c$ suppression with an increase of $\rho_0$ through the Co, Ni and Zn substitution are different from one another. The rates of the $T_c$ suppression are hardly explained in terms of the pair-breaking effect due to the potential scattering by nonmagnetic impurities calculated based on the AG theory, even if errors in the estimation of $n$, $m^*$ and $\rho_0$ were taken into account. The large and small suppression of $T_c$ through the Co and Ni substitution, respectively, have been interpreted as being due to the electron doping. The suppression of $T_c$ through the Zn substitution without electron doping has been found to be very small. In conclusion, the present results are not explained in terms of the $S_\pm$-wave symmetry of the superconducting gap but suggest the presence of the $S_{++}$-wave symmetry in $FeSe_{1-x}Te_x$.

**Acknowledgments**

We would like to thank T. Mizokawa, M. Sato and H. Kontani for helpful discussion. We are grateful to M. Ishikuro of the Institute for Materials Research (IMR), Tohoku University, for his aid in the ICP-OES analysis. This work was supported by a Grant-in-Aid for Scientific Research from the Japan Society for the Promotion of Science.

Figure captions

Fig. 1. (Color online) Dependence on $y$ of the lattice parameters $a$ and $c$ of $Fe_{1-y}M_ySe_{0.3}Te_{0.7}$ (M = Co, Ni, Zn) single crystals.

Fig. 2. (Color online) Temperature dependence of the magnetic susceptibility, $\chi$, in a magnetic field of 1 T parallel to the $c$-axis for $FeSe_{0.3}Te_{0.7}$ and $Fe_{0.95}M_{0.05}Se_{0.3}Te_{0.7}$ (M = Co, Ni, Zn) single crystals.

Fig. 3. (Color online) Temperature dependence of the Hall coefficient, $R_H$, of $FeSe_{0.3}Te_{0.7}$ and $Fe_{0.95}M_{0.05}Se_{0.3}Te_{0.7}$ (M = Co, Ni, Zn) single crystals.

Fig. 4. (Color online) Temperature dependence of the magnetic susceptibility, $\chi$, in a magnetic field of 10 G parallel to the $c$-axis on warming after zero-field cooling for (a) $Fe_{1-y}Co_ySe_{0.3}Te_{0.7}$, (b) $Fe_{1-y}Ni_ySe_{0.3}Te_{0.7}$ and (c) $Fe_{1-y}Zn_ySe_{0.3}Te_{0.7}$ single crystals.

Fig. 5. (Color online) Temperature dependence of the electrical resistivity along the $ab$-plane, $\rho_{ab}$, of several pieces of $Fe_{1-y}M_ySe_{0.3}Te_{0.7}$ (M = Co, Ni, Zn) single crystals. Insets show enlarged plots of $\rho_{ab}$ at low temperatures.

Fig. 6. (Color online) Dependence on $y$ of $T_c$ estimated from $\rho_{ab}$ (closed symbols) and $\chi$ (open symbols) measurements for $Fe_{1-y}M_ySe_{0.3}Te_{0.7}$ (M = Co, Ni, Zn) single crystals. Bars indicate the temperatures where $\rho_{ab}$ drops 90 and 10 % of the normal-state resistivity. Arrows indicate that crystals are not superconducting above 2 K.

Fig. 7. (Color online) Variation of $T_c$ with (a) the minimum value of the residual resistivity, $\rho_0^{min}$, and (b) the maximum value of the residual resistivity, $\rho_0^{max}$, for $Fe_{1-y}M_ySe_{0.3}Te_{0.7}$ (M = Co, Ni, Zn) single crystals. The $T_c$ is defined as the midpoint of the superconducting transition curve in the $\rho_{ab}$ vs. $T$ plot. Bars indicate the temperatures where $\rho_{ab}$ drops 90 and 10 % of the normal-state resistivity. Shadow areas are expected ones in a $S_{\pm}$-wave superconductor calculated using Eq. (2) and taking into account errors in the estimation of the carrier concentration, $n$, and the effective mass, $m^*$.

Table I.  Chemical compositions determined by ICP-OES of as-grown single-crystals of $Fe_{1-y}M_ySe_{0.3}Te_{0.7}$ (M = Co, Ni).

| nominal composition | composition (ICP-OES) | |
|---|---|---|
| Fe : M : Se : Te | Fe : Co : Se : Te | Fe : Ni : Se : Te |
| 1.02 : 0.01 : 0.30 : 0.70 | 1.061 : 0.010 : 0.310 : 0.690 | 1.076 : 0.011 : 0.297 : 0.703 |
| 1.01 : 0.02 : 0.30 : 0.70 | 1.067 : 0.021 : 0.308 : 0.692 | 1.067 : 0.021 : 0.310 : 0.690 |
| 1.00 : 0.03 : 0.30 : 0.70 | 1.066 : 0.031 : 0.312 : 0.688 | 1.061 : 0.035 : 0.311 : 0.689 |
| 0.98 : 0.05 : 0.30 : 0.70 | 1.049 : 0.051 : 0.314 : 0.686 | 1.036 : 0.055 : 0.315 : 0.685 |

Fig. 1.

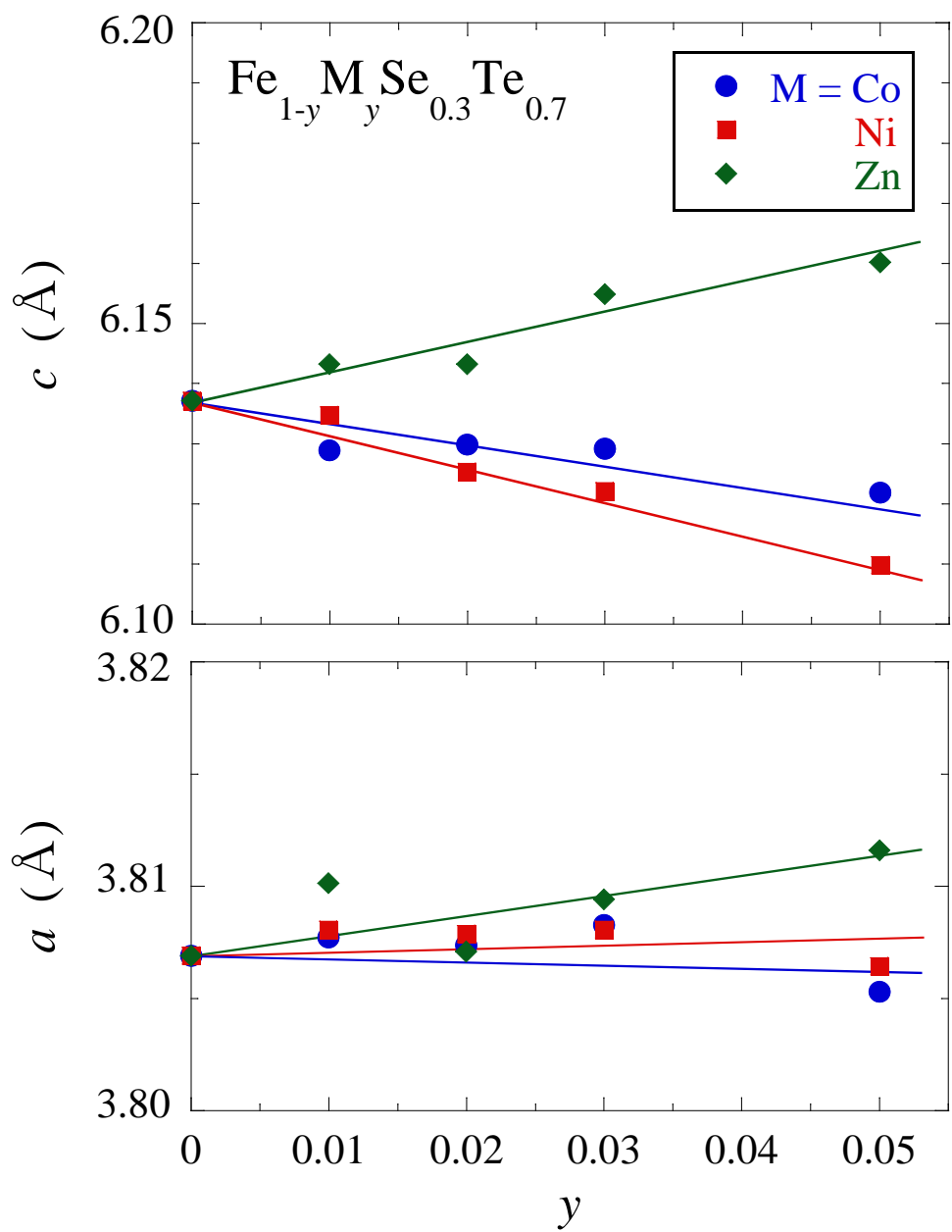

Fig. 2.

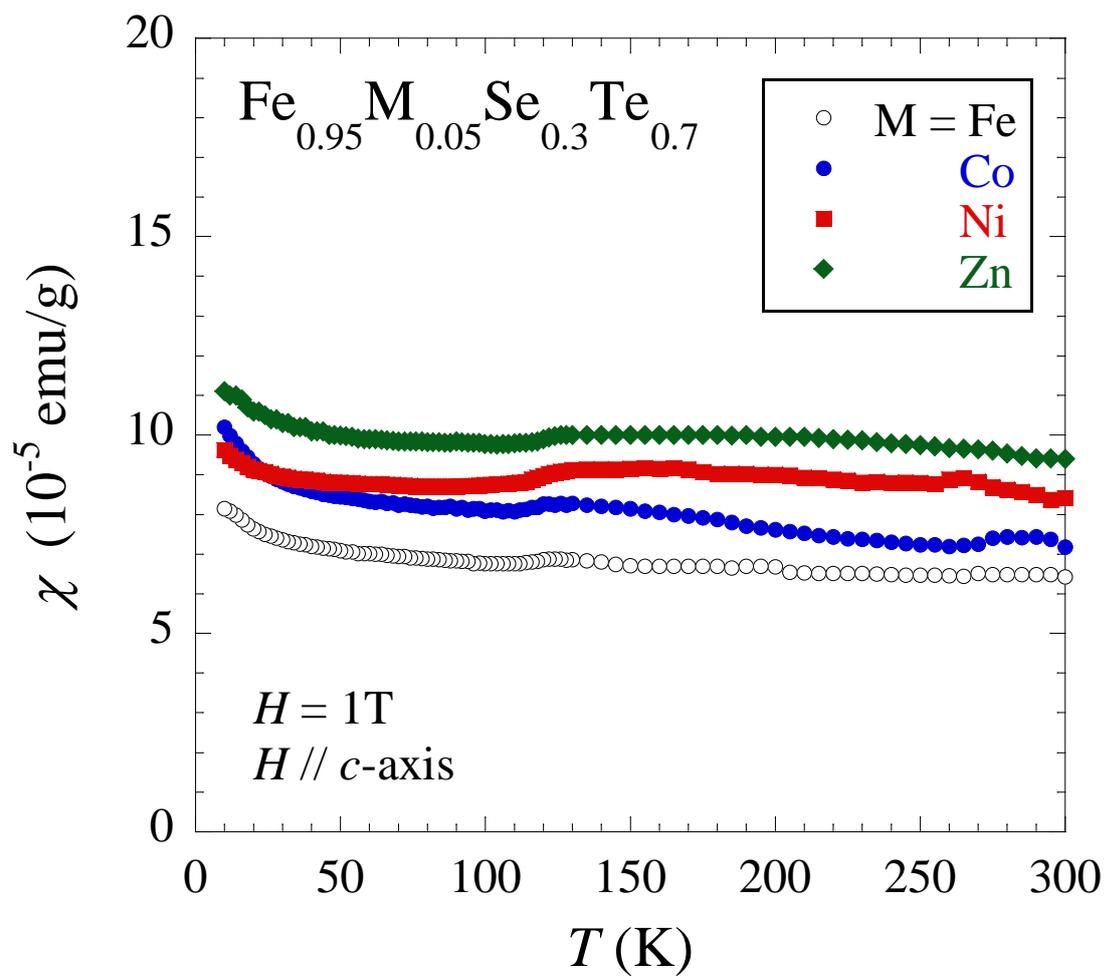

Fig. 3.

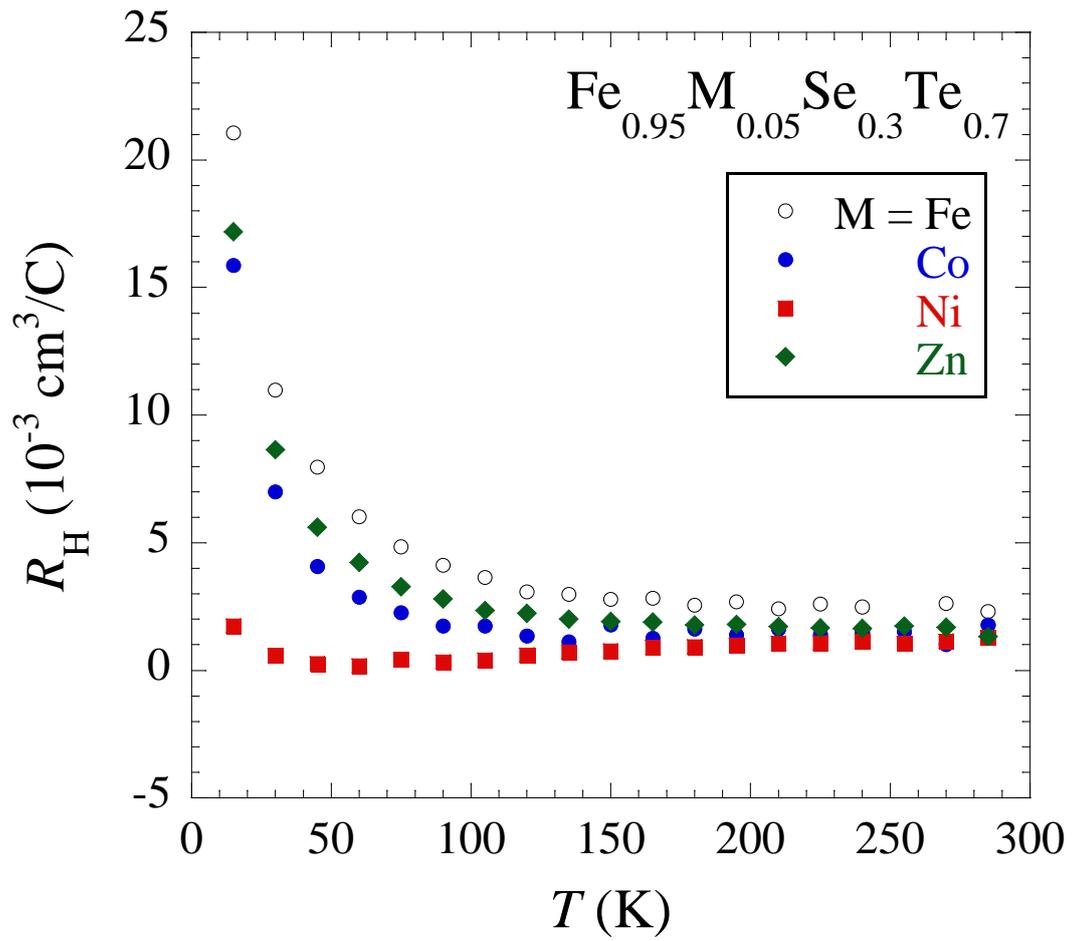

Fig. 4.

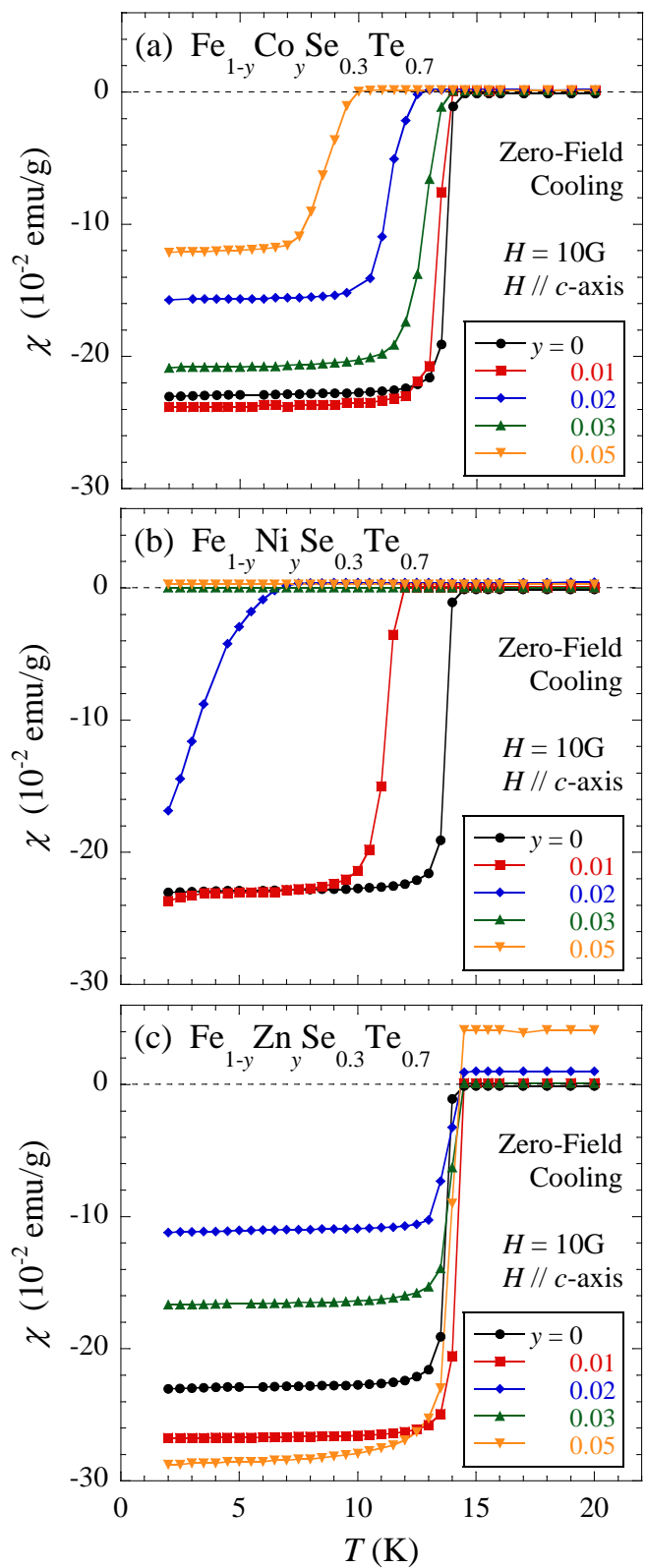

Fig. 5.

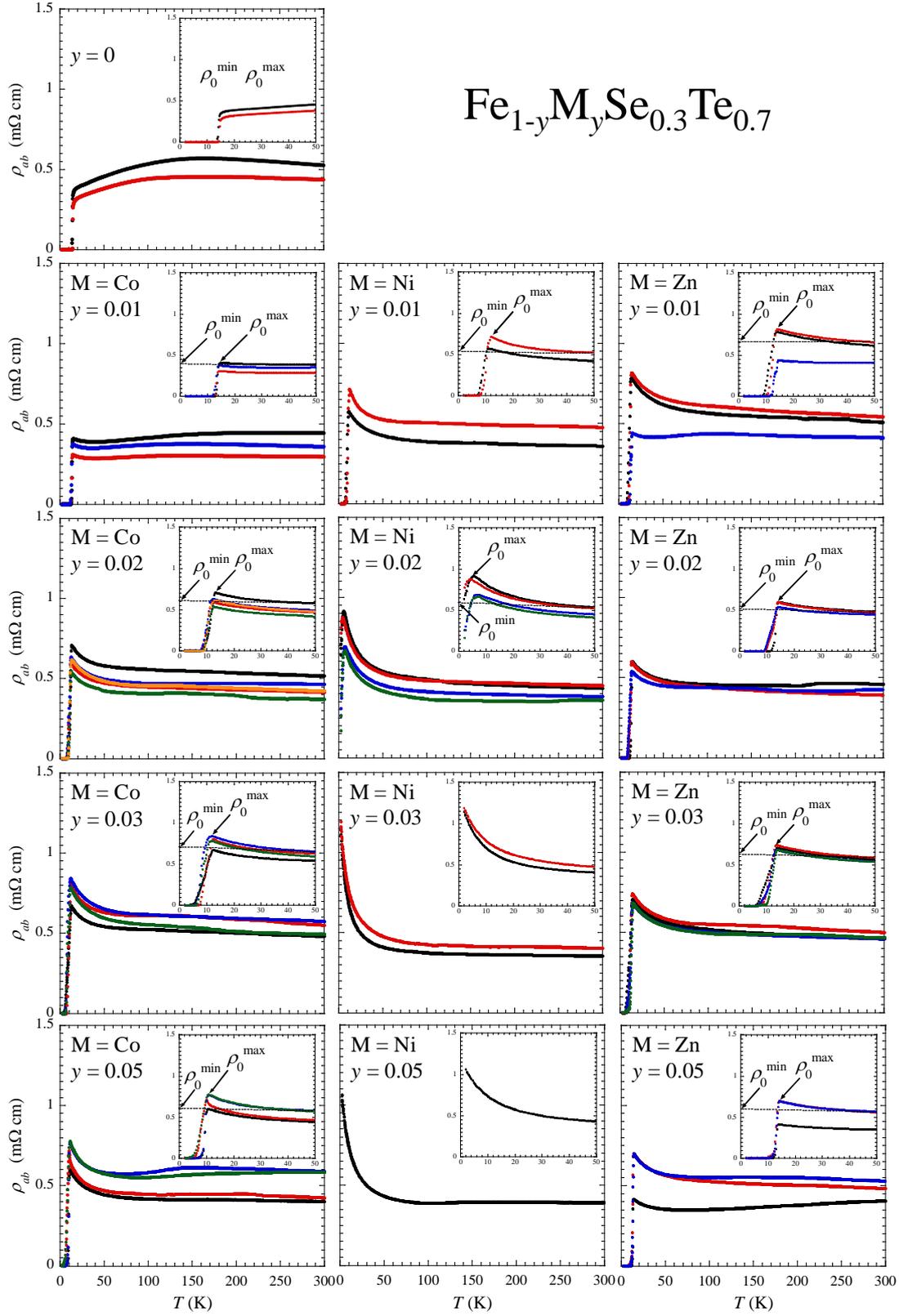

Fig. 6.

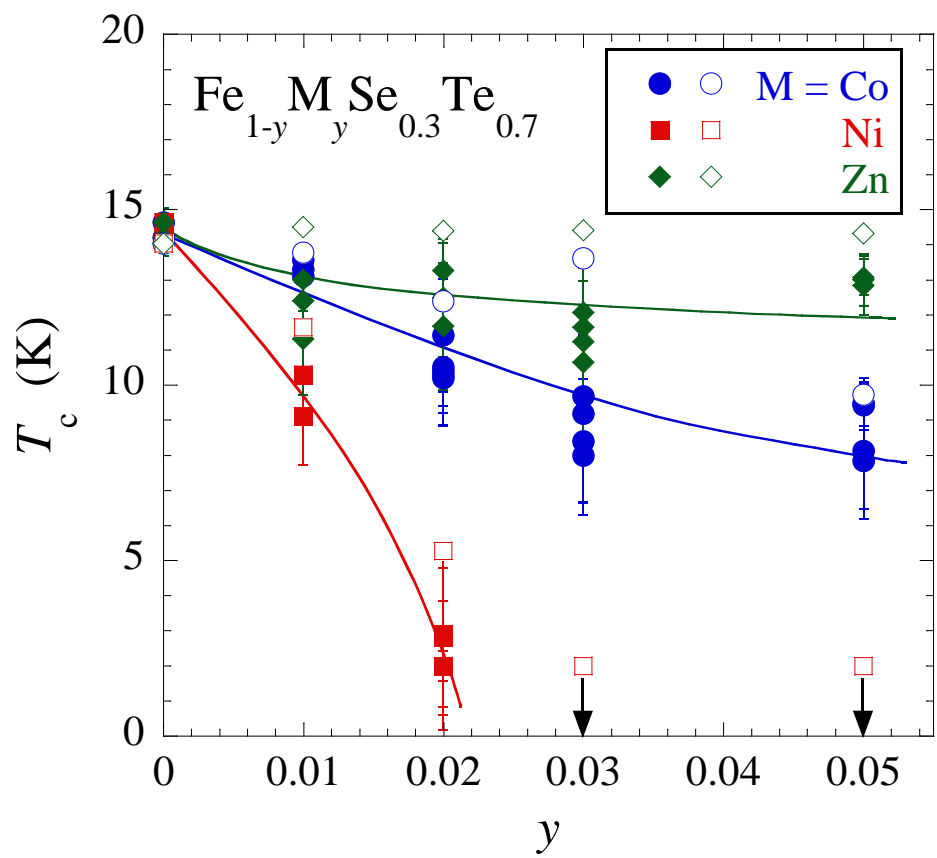

Fig. 7.

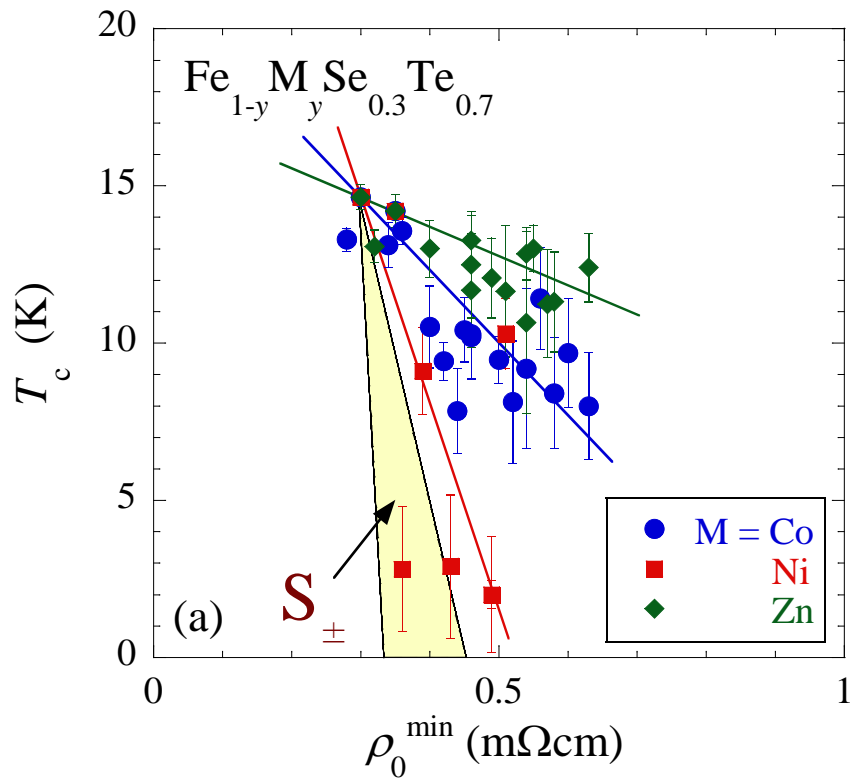

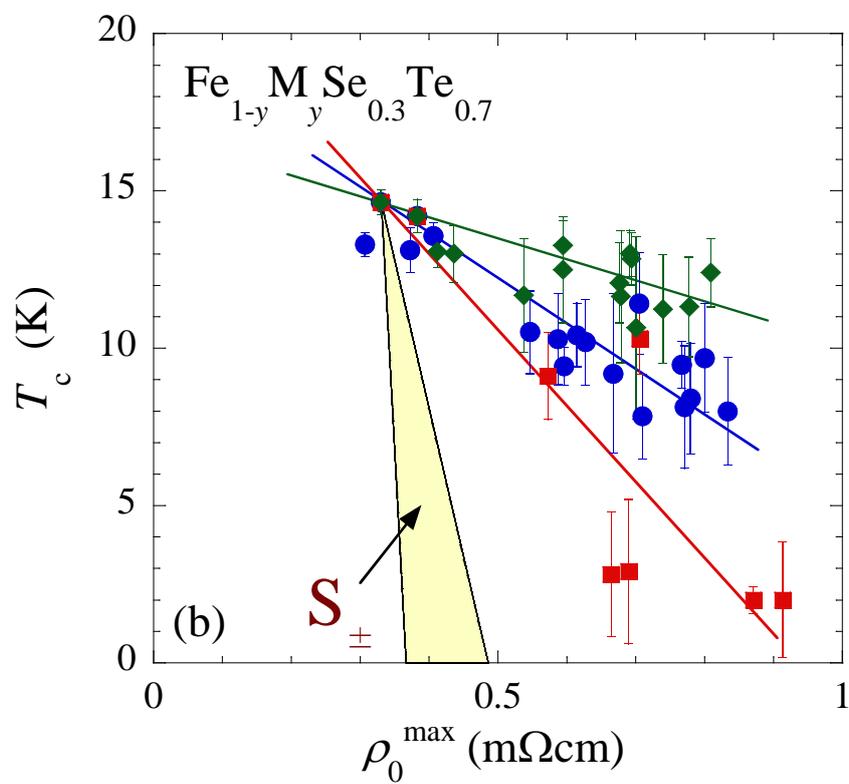